\documentclass[fleqn,12pt,twoside]{article}
\usepackage{espcrc1}

\usepackage{epsfig}

\newcommand{\bm}[1]{\mbox{\boldmath $#1$}}
\newcommand{\st}{{\scriptscriptstyle T}}
\newcommand{\xbj}{x_{\scriptscriptstyle B}}
\newcommand{\bpt}{\bm p}
\newcommand{\bkt}{\bm k_{_T}}

\newcommand{\ba}{\begin{eqnarray}}
\newcommand{\ea}{\end{eqnarray}}
\newcommand{\beq}{\begin{equation}}
\newcommand{\eeq}{\end{equation}}

\newcommand{\slsh}[1]{\mbox{$\not\! #1$}}

\newcommand{\psibar}{\overline{\psi}}
\newcommand{\la}{\langle}
\newcommand{\ra}{\rangle}
\newcommand{\amp}[1]{\la #1 \ra}
\newcommand{\twoamp}[1]{\la \! \la \, #1 \, \ra \! \ra}

\newcommand{\text}[1]{\mbox{$\rm{ #1}$}}
\newcommand{\textsmall}[1]{\mbox{${\scriptstyle \rm{ #1}}$}}

\newcommand{\AmS}{{\protect\the\textfont2
  A\kern-.1667em\lower.5ex\hbox{M}\kern-.125emS}}

\title{\mbox{}\\[-16 mm]
Mapping the Transverse Nucleon Spin\footnote{Invited talk at the
European Workshop on the QCD Structure of the Nucleon (QCD-N'02), Ferrara,
Italy, April 3-6, 2002}}

\author{Dani\"{e}l Boer\address{Department of Physics and Astronomy, 
Vrije Universiteit\\ 
De Boelelaan 1081, 1081 HV Amsterdam, The Netherlands}}
       
\begin{document}

\maketitle

\begin{abstract}
The transverse nucleon spin can be transferred to the quarks and gluons in
several ways. In the factorizing, hard scattering processes to be
considered, these are parameterized at leading twist by the transversity 
distribution function and at next-to-leading twist by quark-gluon correlation 
functions. The latter enter the
description of the structure function $g_2$ and possibly of single spin 
asymmetries. It is discussed what is 
known about these functions and what are the remaining open issues. 
\end{abstract}

\section{Introduction}

Large single transverse spin asymmetries have been observed experimentally
in the process $p \, p^{\uparrow} \rightarrow \pi \, X$ \cite{Adams}. However,
the question of how the direction of the pions is correlated with the 
transverse spin direction of the nucleon has not been answered yet. Many 
theoretical studies have been devoted to the possible origin(s) of 
such asymmetries. One clean approach is to consider the process in a 
kinematical
region where factorization applies and hence, where 
a description in terms of 
quark and gluon correlation functions is valid. What are the 
possibilities in such a description will be the main subject here. 

\begin{figure}[htb]
\begin{center}
\begin{minipage}{5.5cm}
\begin{center}
\mbox{}\\[5 mm]
\epsfxsize=5.5cm \epsfbox{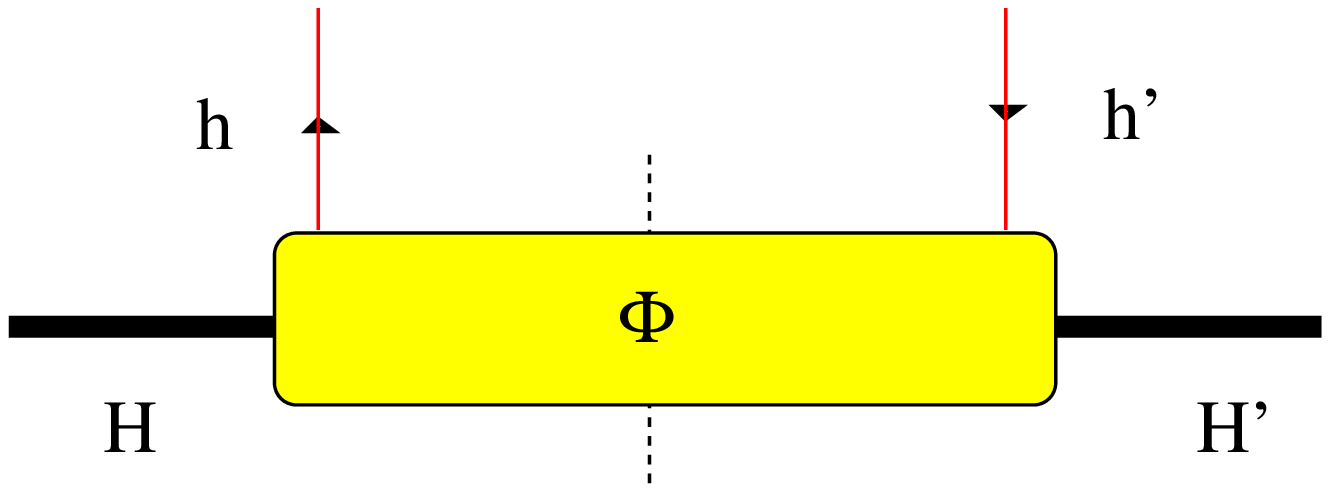}
\mbox{}\\[-4 mm]
\mbox{(a)}
\end{center}
\end{minipage}
\begin{minipage}{10cm}
\begin{center}
\epsfxsize=9.5cm \epsfbox{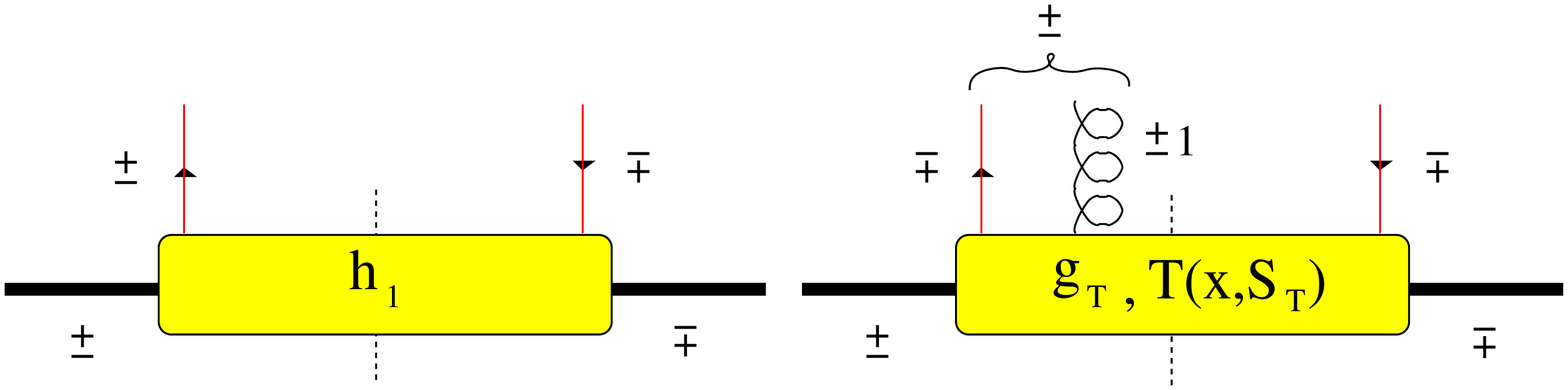}
\mbox{(b)}
\end{center}
\end{minipage}
\vspace{-2 mm}
\caption{\label{helicitylabels} (a) The helicity dependent amplitude $\Phi = 
(H,h;H',h')$; (b) The helicity dependence of $h_1, \, g_T$ and
$T(x,S_T)$.}
\end{center}
\vspace{-4 mm}
\end{figure}

A transverse spin state is an off-diagonal state in the helicity 
basis. For instance, consider a state of transverse polarization in the
(conventional) $\pm\, {\hat x}$ direction:
\beq
\begin{array}{ccc}
| \! \uparrow \ra & = & 
\left[\, | \bm{+} \ra + | \bm{-} \ra\, \right]/\sqrt{2} \\[2 mm]
| \! \downarrow \ra & = & \left[\, | \bm{+} \ra - | \bm{-} \ra\, 
\right]/\sqrt{2}
\end{array} \ \Bigg\} \quad \Longrightarrow \quad
| \! \uparrow \ra \la \uparrow \! |  - | \! \downarrow \ra \la 
\downarrow \! | = | {\bm{+}} \ra \la {\bm{-}} | \, {+} \, 
| {\bm{-}} \ra \la {\bm{+}} |.  
\eeq
Clearly the r.h.s.\ is a helicity-flip density matrix element. 
In factorized hard 
scattering processes one encounters helicity dependent amplitudes 
$\Phi = (H,h;H',h')$ (Fig.\ \ref{helicitylabels}a), where $H, H'$ are nucleon
and $h,h'$ are quark helicity labels. 
In this notation the matrix element $h_1  \equiv  (+,+;\, -,-)$ (called 
transversity) signals helicity flip. The transversity 
distribution function $h_1^q$ 
\cite{Ralst-S-79} (also commonly denoted by $\delta q$)
is a measure of how much of the transverse spin of a polarized nucleon is 
transferred to its quarks. In other words, it is the density of transversely 
polarized quarks inside a transversely polarized nucleon and is 
a function of the lightcone momentum fraction $x$ of a quark inside the 
nucleon. 

There is no diagram Fig.\ \ref{helicitylabels}a for gluons, at least
at leading 
order where the gluons carry $\pm 1$ helicity. Gluons play a subleading role 
(either $h$ or $h'$ should be helicity 0), i.e.\ 
they enter the cross section  
suppressed by one or more powers of the hard scale. Therefore the 
transversity distribution does not mix 
with a gluon distribution function. Under evolution, gluons which are radiated
off only carry away momentum, but do not affect transverse
quark spin (perturbatively gluons couple in a helicity conserving way). 

Although gluons play a subleading role, they may be crucial for the single
spin asymmetries mentioned above. At subleading order one has two types
of gluon dependent matrix elements that play a role: the aforementioned
two-gluon correlation function and a quark-gluon-quark correlation. These two
matrix elements will in fact mix under evolution and here we will focus 
attention only on the latter. In Fig.\ \ref{helicitylabels}b 
its helicity structure is shown and compared to the transversity function. 
In the quark-gluon correlation matrix elements, $g_T$ and
$T(x,S_T)$, the gluon balances the helicity flip of the nucleon state, such
that there is no helicity flip required of the quark states. The difference
between $g_T$ and $T(x,S_T)$ is that in the latter matrix element the
gluon has zero momentum fraction, which occurs only in the description of 
single spin asymmetries. 

\section{Transversity}

\noindent
As said the transversity distribution function 
is a helicity flip (so-called chiral-odd) amplitude. Observables involving 
transversity should therefore be (helicity flip)$^2$. This is the reason why
$h_1$ cannot be measured in inclusive Deep Inelastic Scattering (DIS) 
$e \, p \to e' \, X$; it enters the cross section suppressed by a 
factor of order $m_q/Q$, where $m_q$ is the mass of the struck quark and
$Q$ is the invariant mass of the virtual photon that probes the nucleon, for
which we take a proton from now on.
A further complication is that in charged current exchange processes 
chiral-odd functions like $h_1$ cannot be accessed. 

To measure transversity there are essentially 
two options: single or double transverse spin asymmetries in
(semi-inclusive, neutral current) $ep$ and $pp$ processes. 
Few such experiments have been performed to date and 
no (undisputed) experimental information on $h_1$ is available thus far. 
But a number of future experiments (e.g.\ COMPASS, HERMES, RHIC) are 
expected to provide detailed information. 

\subsection{Transversity asymmetries}

\noindent
The Drell-Yan process of two colliding transversely polarized hadrons
producing a lepton pair was originally thought to be the best way to access
the transversity distribution, for instance at RHIC.
This double transverse spin asymmetry $A_{TT}^{DY}$ is proportional to
$h_1^a(x_1) \; \overline h{}_1^a(x_2)$ \cite{Ralst-S-79}.
The problem is that $h_1$ for antiquarks  
inside a proton ($h_1^{\bar a} = \overline h{}_1^a$) is presumably much 
smaller than for quarks and the asymmetry is not expected to be large. 
In fact, by using Soffer's inequality ($|h_1(x)| \leq {\scriptstyle
\frac{1}{2}} \left[ f_1(x) + g_1(x) \right]$), $A_{TT}^{DY}$ has been shown
\cite{Martin} to be 
small at RHIC, probably 
just beyond the experimental reach (but a future luminoscity 
upgrade will be very promising). 

Also the double transverse spin asymmetry in jet production \cite{Jaffe-Saito}
(directly proportional to $h_1^a(x_1) \; h_1^a(x_2)$ at high $p_{_T}$) 
poses experimental problems due to small cross sections and asymmetries. 
In Ref.\ \cite{SofStratVog} it is demonstrated 
that for RHIC statistics is not the problem, but that the
systematic errors need to be under extremely good control. 

Another possible way to access the transversity distribution function via a 
double transverse spin asymmetry, involves the 
transversity {\em fragmentation\/} function $H_1$. It
measures the probability of
$q^\uparrow \to h^\uparrow + X$, where $h$ is a spin-1/2 hadron, 
for instance a 
$\Lambda$ hyperon.
The double transverse spin asymmetry $D_{NN}$ --the transverse
polarization transfer-- involving both $h_1$ and 
$H_1$, occurs in the processes $e \; p^\uparrow \to e' \;  
\Lambda^\uparrow \; X$ and $p \; p^\uparrow \to 
\Lambda^\uparrow \; X$ \cite{DeFlorianStratVog}. 
The latter observable has been measured by E704 \cite{BravarDNN} and found to
be sizable, but a conclusion about $h_1$ cannot be drawn due to the
low $p_{_T}$ range, which prohibits the use of a factorized 
expression for the cross section. Furthermore, $H_1$ is also 
unknown and although it could be extracted from 
$e^+ \: e^- \to \Lambda^\uparrow \; 
\overline{\Lambda}{}^\uparrow \: X$, this also poses quite a challenge.

In short, double transverse 
spin asymmetries do not seem promising for extracting the 
transversity distribution in the near future. This leaves the  
single spin asymmetry (SSA) options, which all exploit fragmentation
functions of some sort. Three options have been considered: 
1) measuring the transverse momentum of the final state 
hadron compared to the jet axis, exploiting the so-called Collins effect; 
2) producing final state hadrons with 
higher spin, e.g.\ $\rho$ or its decay product: a $\pi^+ \pi^-$ pair 
(related to the interference fragmentation functions);  
3) higher twist asymmetries which are suppressed by inverse
powers of the hard scale $Q$. The third option will not be discussed here.

\section{Collins effect asymmetries}

\noindent
The Collins effect refers to a nonzero correlation between 
the transverse spin $\bm s_{_T}$ of a fragmenting quark and the distribution 
of produced hadrons. More specifically, the transversely polarized quark  
fragments into particles (with nonzero transverse momentum $\bkt$) distributed
with a 
$\bkt \times \bm s_{_T}$ angular dependence around the jet axis or, 
equivalently, the quark momentum, see Fig.~1 of Ref.\
\cite{DB-DIS01}. The Collins
effect will be denoted by a fragmentation function $H_1^\perp(z,\bkt)$ and
if nonzero, it can lead to SSA in 
$e \, p^\uparrow \to e' \pi \, X$ and $p \, p^\uparrow \to \pi \, X$. 
There is some experimental hint in the SMC data \cite{Bravar} 
(at the 2$\sigma$ level) that the Collins 
effect (and hence transversity) is indeed nonzero. 
Also, HERMES pion asymmetry ($A_{UL}$) data \cite{HERMES}  
and the left-right asymmetries in $p \, p^\uparrow \to \pi \, X$ \cite{Adams} 
can (at least partially) be described in terms of
the Collins effect (see e.g.\ \cite{Efremov-Goeke,Boglione}). 
No consensus about the interpretation of these data has been reached though.
Another open question is how the Collins effect works precisely, i.e.\ how 
the transverse spin of quarks is transformed into orbital angular momentum 
during the (nonperturbative) fragmentation process. In any case, 
a correlation between the 
transverse spin of the fragmenting quark and the transverse momentum of a 
hadron in the jet is allowed by the symmetries. 

\subsection{Collins effect in semi-inclusive DIS}

\noindent
Collins considered \cite{Collins-93} semi-inclusive DIS (SIDIS) 
$e \, p^\uparrow
\to e' \, \pi \, X$, where the spin of the 
proton is orthogonal to the 
direction of the virtual photon $\gamma^*$ and one observes the pion
transverse momentum $\bm P^{\pi}_{\perp}$, which has 
an angle $\phi^e_\pi$ compared to the lepton scattering plane. 
He showed that the cross section for this process can 
have an asymmetry 
that is proportional to the transversity function: $A_{T} \propto  
\sin(\phi^e_\pi + \phi^e_{_S}) \, |\bm S_{T}^{}| \, h_1 \, H_1^\perp$.
Since the functions $h_1$ and $H_1^\perp$ are chiral-odd and the asymmetry on
the parton level is $\hat{a}_{TT}$, this asymmetry depends on the orientation
of the lepton scattering plane, i.e.\ if one integrates over
the direction of the back-scattered electron $e'$, then the asymmetry would
vanish. 

To discuss this SSA further, we will focus on the observable 
(cf.\ Refs.\ \cite{Boer-Mulders-98,DB-DIS01,DB-Trento})
\beq
{\cal O} \equiv \frac{\big\langle 
\sin( \phi_{_C} ) \, |\bm P^{\pi}_{\perp}|  
\big\rangle}{{\scriptstyle 
\left[4\pi\,\alpha^2\,s/Q^4\right]}M_\pi} = 
\vert \bm S_{T}\vert\,{\scriptstyle (1-y)} 
\sum_{a,\bar a} e_a^2
\,x\,h_{1}^{a}(x) z H_1^{\perp (1) a}(z),
\label{observableO}
\eeq
where $\phi_{_C} = \phi^e_\pi+\phi^e_{_S}$ and $H_1^{\perp (1)}(z) =  
\int d^2 \bkt \, \bkt^2 \, H_1^{\perp}(z,\bkt^2)/(2 z^2 M_\pi^2)$.
At present all phenomenological studies of the Collins effect are performed 
using tree level expressions like Eq.\ (\ref{observableO}). 
But the leading order (LO) evolution equations 
are known for both $h_1$ (NLO even) and $H_1^{\perp (1)}$ 
(at least in the large $N_c$ limit \cite{Henneman}). The LO $Q^2$ 
behavior of the observable ${\cal O}$ arises solely from the LO 
evolution of $h_1$ and $H_1^{\perp (1)}$. This is however 
a nontrivial result, 
since this semi-inclusive process is not a case where collinear factorization
applies. In the differential cross section $d\sigma/d^2\bm P^{\pi}_{\perp}$ 
itself, beyond tree level soft gluon corrections do not cancel, such that 
Sudakov 
factors need to be taken into account and a more complicated factorization 
theorem applies \cite{CS-81,DB-01}. In fact, the observable ${\cal O}$ 
(Eq.\ (\ref{observableO})) is the only $|\bm P^{\pi}_{\perp}|$-moment 
of the Collins asymmetry in the cross section, that is not sensitive 
to Sudakov factors. This observable is therefore better 
suited for a $Q^2$ dependent analysis 
than the full $|\bm P^{\pi}_{\perp}|$-dependent asymmetry. 

Another point to note here is that the Collins effect asymmetry is not power
suppressed. Since it is not down by $1/Q$, it is sometimes referred to as
leading twist, but strictly speaking the Collins function is not related
to local operators of twist-2 exclusively (rather one is using Jaffe's 
``working redefinition of twist'' \cite{Jaffe}
to indicate the leading power in $1/Q$ at
which the contribution appears in the cross section). The reason why the
asymmetry is not suppressed by $1/Q$ is that one is dealing with a
multi-scale process: the hadronic scale $M_\pi$, 
the transverse momentum $|\bm P^{\pi}_{\perp}|$ of the
pion compared to the virtual photon (which itself 
is in the lepton scattering plane)
and the hard scale $Q$. The Collins effect is a term 
$\slsh{k}_{_T} \slsh{P}{}_{\pi} H_1^\perp(z,\bkt^2)/M_\pi^2$ in the
fragmentation 
correlation function $\Delta(z,\bkt)$ \cite{Coll-S-82}, which indicates  
that the explicit factor of $k_{_T}$ is not suppressed by an inverse power 
of the hard scale $Q$, but rather is compensated by a hadronic scale 
($M_\pi$ by definition). 

The evolution of $H_1^{\perp(1)}$ is quite involved since it is not of
definite twist. One must perform a full   
twist-3 evolution calculation \cite{Belitsky}. Luckily, 
in the large $N_c$ limit the evolution simplifies (it becomes DGLAP-like 
\cite{Henneman}) and should be sufficient for initial comparisons of data 
which are obtained at different energies. 

The effect of Sudakov factors is to 
decrease the magnitude of the asymmetry and to broaden it as a function of 
transverse momentum. This decrease can be quite substantial at high energies, 
estimated to be 
roughly $1/\sqrt{Q}$ per Collins function when compared to tree level
\cite{DB-01}. As said, to avoid such Sudakov suppression one can consider the
particular moment in Eq.\ (\ref{observableO}). But one does need to keep in
mind that the average transverse momentum of the pion, 
$\amp{{|\bm P^{\pi}_{\perp}|}}$, is a function of $Q$, 
governed by Sudakov factors. 

\subsection{Collins effect in $e^+ \, e^- \to \pi^+ \, \pi^- \, X$} 

\noindent
In order to obtain the Collins function itself,
one can measure a $\cos(2\phi)$
asymmetry in $e^+ \, e^- \to \pi^+ \, \pi^- \, X$, that has a contribution 
which is proportional to the Collins function squared \cite{BJM-97} (at equal
momentum fractions). 
A first indication of such a nonzero (but small) asymmetry comes from a 
preliminary analysis \cite{EST} of the LEP1 data (DELPHI). 
A similar study of off-resonance data from 
the B-factory BELLE at KEK, is under way \cite{Ogawa}.

Also for this Collins effect observable the tree level 
asymmetry expression is not sufficient when 
results from different experiments
are to be compared. Beyond tree level Sudakov factors 
need to be included. Since the Collins function enters twice in this asymmetry
the suppression is estimated to be of the order $1/Q$ $(= 1/\sqrt{s})$ 
\cite{DB-01}, which may well be the reason the {DELPHI} 
data indicated a small
asymmetry. Therefore, this Collins 
effect observable is best studied with two jet events at lower $\sqrt{s}$
(a requirement satisfied by {BELLE}, which operates on and just below 
the $\Upsilon(4S)$, i.e.\ around 10.5 GeV). 

Nevertheless, the extraction of the Collins function from this asymmetry is 
not straightforward, since there is asymmetric background from hard gluon 
radiation (when $Q_T \sim Q$) and from weak decays. The former enters the
$Q_T$ dependent asymmetry proportional to $\alpha_s Q_T^2/Q^2$, which 
at lower values of $Q^2$ need not be small. This
contribution could be neglected at LEP energies 
\cite{BJM-97}. Luckily it is calculable and so is 
the background from weak decays,
e.g.\ $e^+ e^- \to \tau^+ \tau^- \to \pi^+ \pi^- X$. 

As in the case of the Collins asymmetry in SIDIS, there is one particular
$Q_T$ moment of the asymmetry that is not sensitive to Sudakov factors, 
namely the first $Q_T^2$ moment: $\int dQ_T^2 Q_T^2 d\sigma/dQ_T^2$.
Unfortunately, it is mostly 
sensitive to the high $Q_T^2$ $(\sim Q^2)$ hard gluon radiation. 
This contribution could in principle be cut off
by imposing a maximum $Q_T$ cut, but this introduces a further source of 
uncertainty. It may therefore be better to calculate its contribution 
using the known ordinary fragmentation functions $D_1^\pi$.  

The main conclusion we can draw about this kind of 
Collins effect asymmetries is that they are not like ordinary leading
twist asymmetries and special care must be taken beyond tree level regarding
Sudakov factors. 

\section{Interference fragmentation functions}

\noindent
Apart from the Collins effect, there may be a 
correlation between the transverse spin of the fragmenting quark and the 
orientation of a $\pi^+ \, \pi^-$ pair inside the jet
\cite{Ji-94,ColHepLad,JJT}, which can also be parameterized by a chiral odd
function. Concretely, Jaffe, Jin and Tang \cite{JJT} 
considered the two pion final state
$\vert (\pi^+ \, \pi^-)_{{\textsmall{out}}} X \rangle $ ($\pi^+, \pi^-$ 
belong to the same jet) and assumed a dependence on the strong phase 
shifts of the $\pi^+ \, \pi^-$ system. The interference between different 
partial waves then gives rise to a nonzero chiral-odd fragmentation function, 
called the interference fragmentation function (IFF). 
The IFF would lead to single spin asymmetries in 
$ e \, p^\uparrow \rightarrow e' \, (\pi^+ \, \pi^-) \, X$ and 
$p\, p^\uparrow \to (\pi^+ \, \pi^-) \, X$, both proportional to $h_1$. 
The SSA expression\footnote{In Refs.\ \cite{DB-DIS01,DB-Trento} 
the asymmetry Eq.\ (\ref{JJTasym}) was erroneously written as $\cos(
\phi_{S_T}^e + \phi_{R_T}^e)$.}
$e \, p^\uparrow \rightarrow e' \, (\pi^+ \, \pi^-) \, X$ in terms of the IFF
$\delta\hat q_I^{}(z)$ is \cite{JJT} 
\beq
\big\langle \sin( \phi_{S_T}^e + \phi_{R_T}^e) 
\big\rangle \propto  F    
\vert \bm S_{T} \vert \vert 
\bm R_{T}\vert h_{1}(x) \delta\hat q_I^{}(z), 
\label{JJTasym}
\eeq
where $z=z^+ + z^-$; $\bm R_{T} 
= (z^+ \bm k^- - z^- \bm k^+)/z$; $F = F(m^2) = \sin \delta_0 \sin \delta_1
\sin(\delta_0-\delta_1)$, where $\delta_0, \delta_1$ are the $\ell = 0, 1$ 
phase shifts and $m^2$ is the $\pi^+ \pi^-$ invariant 
mass.
Note again the dependence on the orientation of the lepton scattering plane. 
Also note the implicit assumption of factorization of $z$ and $m^2$ 
dependences, which leads to the prediction that on the $\rho$
resonance the asymmetry is zero (according to the
experimentally determined phase shifts). More general $z, m^2$ 
dependences have been considered \cite{Bianconi}. 

Unlike the Collins effect asymmetries, Eq.\ 
(\ref{JJTasym}) is based on a collinear factorization 
theorem (soft gluon contributions cancel, no Sudakov factors appear). 
This makes an analysis beyond tree level conceptually straightforward. 
The evolution of $\delta \hat{q}_I(z)$ equals that of $H_1(z)$ (known to 
NLO \cite{Stratmann-Vogelsang-01}) and a NLO analysis is feasible (cf.\ also
Ref.\ \cite{Contog}). 

For the extraction of 
the interference fragmentation functions themselves one can study a 
$\cos(\phi_{R_{1T}}^e + \phi_{R_{2T}}^e)$ asymmetry \cite{Artru-Collins} in   
$e^+ \, e^- \, \to \, (\pi^+ \, \pi^-)_{{\textsmall{jet} \, 1}} \, (\pi^+ \,
\pi^-)_{{\textsmall{jet} \, 2}} \, X$ which is proportional to $(\delta\hat
q_I)^{2}$. This is again possible at {BELLE} and this time there is no 
expected asymmetric background. Combining such a result with for instance  
the single spin asymmetry in $p\, p^\uparrow \to \pi^+ \pi^- \, X$ to be 
measured at RHIC or in $e\, p^\uparrow \to e' \, \pi^+ \pi^- \, X$ at COMPASS
or HERMES, seems --at present-- to be one of the
most realistic ways of obtaining information on the transversity function. 
However, one does need to have an accurate measurement of the ordinary 
fragmentation function of a quark into a $\pi^+ \, \pi^-$ pair, 
$D_1^{\pi^+ \, \pi^-}$.  

\section{Quark-gluon correlations}

Gluons play a subleading role in the transverse spin of the nucleon, but their
effects may be observable nevertheless. But so far not more than a hint 
comes from the measurement of the structure function $g_2$. 
At tree level the structure 
function ${g_2(\xbj,Q^2)}$ is expressed in terms of parton distribution
functions as ${g_2(\xbj,Q^2)} = \sum_{a,\bar{a}} 
e_{a}^2 \left({g_T^a(\xbj)} - {g_1^a(\xbj)} \right)/2$,
where ${g_T}$ is {not} a {\em density\/} however. It enters in 
a {power suppressed} 
azimuthal spin asymmetry in the cross section of polarized DIS 
\beq
\frac{d\sigma (\vec \ell H^\uparrow \rightarrow \ell^\prime X)}
{d\xbj\,dy} \propto
\biggl\{ 
{\scriptstyle \left(\frac{y^2}{2} + 1- y \right)} \, \xbj {f^a_1(\xbj)}
- {\scriptstyle 2 y \sqrt{1-y}}\; {\lambda_e\,\vert \bm{S}_T \vert} \, 
{\cos (\phi_s)}\,{\frac{M}{Q}}\,{\xbj^2\, {g^a_T(\xbj)}}
\biggl\}.
\label{cosgT}
\eeq
The E155 Collaboration at SLAC has successfully measured $g_2$
\cite{E155}. Within errors it
is still consistent with the Wandzura-Wilczek part, which is determined by the
twist-2 function $g_1$. The SLAC data
results in {$d_2^p \equiv 3 \int_0^1 x^2
{\left. g_2(x)\right|_{\textsmall{twist-3}}} dx = 0.0032 \pm 0.0017$}, 
which is barely 2$\sigma$ away from zero. A future demonstration of a nonzero 
${\left. g_2(x)\right|_{\textsmall{twist-3}}}$ would
unambiguously show the role of gluons in the transverse nucleon spin. 
 
The azimuthal angular 
dependence $\cos (\phi_s)$ (the angle $\phi_s$ is between
the transverse spin and the lepton scattering plane) in Eq.\ (\ref{cosgT}) is
the only possible one in fully inclusive DIS $\vec \ell \vec H \rightarrow 
\ell^\prime X$. Christ and Lee \cite{ChristLee} have shown that a 
$\sin (\phi_s)$ dependence would violate time reversal invariance. Since
parity forces such a $\sin (\phi_s)$ asymmetry to be independent of the lepton
polarization, one would be dealing with a SSA in fully inclusive polarized 
DIS. In general, to generate a SSA one needs to have an imaginary
part. For instance, a complex-valued (but hermitian) quark-gluon correlation 
function, parameterized by $g_T$ and $f_T$ (called 
a T-odd distribution function), yields in the cross section a power-suppressed
$\sin (\phi_s)$ single transverse spin asymmetry:
\beq
\frac{d\sigma (\ell H^\uparrow \rightarrow \ell^\prime X)}
{d\xbj\,dy} \propto
\biggl\{ 
{\scriptstyle \left(\frac{y^2}{2} + 1- y \right)} \, \xbj {f^a_1(\xbj)}
- {\scriptstyle 2 y \sqrt{1-y}}\; {\vert \bm{S}_T \vert} \, 
{\sin (\phi_s)}\,{\frac{M}{Q}}\,{\xbj^2\, {f^a_T(\xbj)}}
\biggl\}.
\eeq
As said this would violate time reversal invariance. So far the absence of
such an asymmetry has only been confirmed with percent level precision in 
experiments
performed more than 30 years ago \cite{Chen-68}\footnote{I thank A. Drago for
pointing out these references.}.

To formulate it more precisely, 
T-odd distribution functions are {absent} due to time reversal 
invariance, {\em if\/} the incoming hadron is treated as a {plane-wave 
state} \cite{Collins-93}. This is related to the stable 
nature of the incoming hadron (stable on strong interaction time scales, since
the incoming hadron might be a neutron) and the absence of imaginary parts 
in its correlation functions.

However, a formal loophole has been pointed out by 
Anselmino {\em et al.\/} \cite{AnselminoBDM}. In addition, there may be
other ways of generating an imaginary part if the process is not fully 
inclusive. This is the case for the Efremov-Teryaev and Qiu-Sterman single 
spin asymmetries \cite{ET-85,Qiu-Sterman-90s}. 
Qiu and Sterman considered SSA
in pion and prompt photon production arising from a so-called soft gluon (or
gluonic) pole mechanism. Hammon {\em et al.\/}~\cite{HTS}
applied this to the Drell-Yan (DY) process. Here the contribution to the SSA
comes from a gluon with vanishing lightcone momentum fraction, 
such that the $i\epsilon$ part of a fermion propagator is picked up. 
It yields {an effective $f_T$ function}, which means that the effects 
of the gluonic pole are indistinguishable from those of an $f_T$ \cite{BMT}.  
The only reason that one cannot view it as being equivalent to a T-odd
distribution function is that the gluonic pole does not contribute to fully
inclusive DIS (which is the only exception). 

The Qiu-Sterman matrix element $T(x,S_T)$ has the 
following operator structure  
\ba
{T(x,S_T)} & \propto & \twoamp{\psibar (0) {\Gamma_\alpha} 
{\int d\eta \; F^{+\alpha} (\eta n_-)}\; \psi(\lambda n_-)}.
\ea
The lightcone integral over the field strength reflects the fact that 
the gluon has
vanishing lightcone momentum fraction. Since this quantity is like an
average gluon field strength inside the ordinary unpolarized quark
distribution function, as a first guess one sometimes uses the simplifying 
Ansatz (due to Qiu and Sterman): $T^a(x,S_T) \approx \kappa_a\, \lambda\,
f_1^a(x)$, with $\kappa_u=1=-\kappa_d$, $\kappa_s=0$. Using the parameter 
value $\lambda \sim 100$~MeV obtained from a fit to the pion production SSA,
the SSA in DY then becomes $\left|A_N\right| \sim  0.7 \, 
\lambda/Q$ \cite{DBQiu}. Hence, just below and above the $J/\psi$ the
asymmetry is of the order of a few percent. 
This prediction is very different from for instance the one by
Boros {\it et al.\/}~\cite{Boros-93}, which ranges well above 20\%. RHIC
experiments should be able to test these predictions in the coming years. In
addition, the Collins effect cannot contribute to the SSA in DY, making the
latter observable an even more useful tool to distinguish between different 
mechanisms.

As a final topic, we now look at the almost inclusive DIS process 
$e \, p^\uparrow \to e' \, \text{jet} \, X$, where one observes the transverse
momentum $\bm P^{\textsmall{jet}}_{\perp}$ of the jet. 
Also here the Collins effect (and hence transversity) 
cannot contribute to a SSA. 

Consider the cross section of an unpolarized electron scattering off a
transversely polarized hadron, weighted by a function of the transverse 
momentum of the jet:  
\beq
\langle {W}\rangle_{{U T}}
\equiv \int dz \; d^2\bm P^{\textsmall{jet}}_{\perp}
\ {W}\,\frac{d\sigma^{[{e \, p^\uparrow \rightarrow e' \,
\textsmall{jet} \, X}]}} {dx\,dy\,dz\,d\phi^e\,d\phi^e_{\textsmall{jet}} d|\bm P^{
\textsmall{jet}}_{\perp}|^2},
\eeq
where ${W}$ = 
$W(|\bm P^{\textsmall{jet}}_{\perp}|,\phi_{\textsmall{jet}}^e)$, but again   
restricted to the case $|\bm P^{\textsmall{jet}}_{\perp}|^2 \ll Q^2$.

For ${W=1}$ one would retrieve the $1/Q$ suppressed asymmetry
proportional to $f_T$ \cite{Boer-Mulders-98}:
\beq
\frac{\langle 1
\rangle_{{UT}}}{{\scriptstyle \left[4\pi\,\alpha^2\,s/Q^4\right]}} = -
\sin\phi_S^e \; \vert\bm S_\st\vert
\,{\scriptstyle (2-y)\,\sqrt{1-y}} \ {\frac{M}{Q}} \, \sum_{a,\bar a} 
e_a^2\, x^2\,{f_T^a(x)} \stackrel{\textsmall{T.rev.}}{=} 0,
\eeq
forced to be zero by time reversal invariance. As mentioned, the
gluonic pole mechanism does not generate it either. 
However, if one weights with a power of the {observed} 
transverse momentum one obtains for instance the following {\em 
unsuppressed\/} expression (take simply
$D_1(z)=\delta(1-z)$ in the expression given in \cite{Boer-Mulders-98})
\beq
\frac{\big\langle 
\cos \phi_{\textsmall{jet}}^e \, {|\bm P^{\textsmall{jet}}_{\perp}| /M} 
\big\rangle_{{UT}}}{{\scriptstyle \left[4\pi\,\alpha^2\,s/Q^4\right]}}
= - \sin \phi_S^e \; \vert \bm S_\st\vert\,
{\scriptstyle (1-y+\frac{1}{2}\,y^2)} 
\sum_{a,\bar a} e_a^2 \,x\,{f_{1T}^{\perp (1)a}(x)}.
\label{WLT}
\eeq
This asymmetry depends on  
the {T-odd Sivers function $f_{1T}^{\perp}(x,\bpt_\st^2)$ \cite{Sivers} via
\beq
{f_{1T}^{\perp (1)}(x)} \equiv \int d^2 \bpt_\st \, \frac{\bpt_\st^2}{2 M^2} 
\, {f_{1T}^\perp(x,\bpt_\st^2)}. 
\eeq

There has been much discussion in the literature whether the T-odd
distribution function ${f_{1T}^\perp}$ 
is allowed to be nonzero or not. Collins 
has given a proof that
it should be zero \cite{Collins-93}, but recently he 
pointed out a loophole \cite{Collins-02}. 
Also, Brodsky, Hwang and Schmidt \cite{BHS}
have shown by an explicit model calculation that there can indeed be
a nonzero (and unsuppressed) asymmetry in the process 
$e \, p^\uparrow \to e' \, \text{jet} \, X$. It arises from a gluon having
zero lightcone momentum fraction (but nonzero transverse momentum), such that
the $i \epsilon$ from a fermion propagator is picked up, generating the
imaginary part required for a SSA. It seems to produce a nonzero Sivers 
effect and thus 
is expected to yield an effective ${f_{1T}^\perp }$ in the same
way as $T(x,S_T)$ generates an effective $f_{T}$.  
It also leads to a SSA in $e \, 
p^\uparrow \to e' \, \pi \, X$, which is distinguishable from the Collins
effect: the latter has a dependence $(1-y)\, \sin(\phi^e_{\pi} + \phi^e_{S})$,
whereas the former $(1-y+y^2/2) \sin(\phi^e_{\pi} - \phi^e_{S})$. 
This is understandable, since the Sivers
effect asymmetry has unpolarized quarks in the elementary cross section and
therefore, does not depend on the lepton scattering plane and also its
$y$ dependence is characteristic of unpolarized scattering. 
Thus, recent investigations \cite{BHS,Collins-02} seem to revive the Sivers
effect as a possible origin of single transverse spin asymmetries, showing
once more that the transverse spin of the nucleon harbors highly nontrivial
physics involving the angular momentum of both quarks {\em and\/} gluons.

\section*{Acknowledgments}

\noindent
I thank Mauro Anselmino, Stan Brodsky, John Collins, Anatoli Efremov, 
Matthias Grosse Perdekamp, Dae Sung Hwang, Bob Jaffe, Rainer Jakob, 
Piet Mulders, Jianwei Qiu, Marco Radici, Akio Ogawa, Naohito Saito, 
George Sterman, Oleg Teryaev, Werner Vogelsang for many fruitful 
discussions.  
The research of D.B. has been made possible by a 
fellowship of the Royal Netherlands Academy of Arts and Sciences.

\end{document}